\documentclass[manuscript]{acmart}

\usepackage{rotating}
\usepackage{geometry}
\usepackage{booktabs}
\usepackage{lscape}
\usepackage{adjustbox}
\usepackage{afterpage}

\AtBeginDocument{%
  }

\setcopyright{acmlicensed}
\copyrightyear{2018}
\acmYear{2018}
\acmDOI{XXXXXXX.XXXXXXX}

\acmConference[Conference acronym 'XX]{Make sure to enter the correct
  conference title from your rights confirmation emai}{June 03--05,
  2018}{Woodstock, NY}
\acmISBN{978-1-4503-XXXX-X/18/06}

\begin{document}

\title[Moving Beyond LDA]{Moving Beyond LDA: A Comparison of Unsupervised Topic Modelling Techniques for Qualitative Data Analysis of Online Communities}

\author{Amandeep Kaur}
\orcid{0000-0003-1822-4160}
\affiliation{% 
% \department {Cheriton School of Computer science}
  \institution{Cheriton School of Computer Science, University of Waterloo}
  \city{Waterloo, Ontario}
  \country{Canada}}
\email{a3kaur@uwaterloo.ca}

\author{James R. Wallace}
\orcid{0000-0002-5162-0256}
\affiliation{
\institution{School of Public Health Sciences, University of Waterloo}
  \city{Waterloo, Ontario}
  \country{Canada}}
\email{james.wallace@uwaterloo.ca}

\renewcommand{\shortauthors}{Kaur et al.}

\begin{abstract}
Social media constitutes a rich and influential source of information for qualitative researchers. Although computational techniques like topic modelling assist with managing the volume and diversity of social media content, qualitative researcher's lack of programming expertise creates a significant barrier to their adoption. In this paper we explore how BERTopic, an advanced Large Language Model (LLM)-based topic modelling technique, can support qualitative data analysis of social media. We conducted interviews and hands-on evaluations in which qualitative researchers compared topics from three modelling techniques: LDA, NMF, and BERTopic. BERTopic was favoured by 8 of 12 participants for its ability to provide detailed, coherent clusters for deeper understanding and actionable insights. Participants also prioritised topic relevance, logical organisation, and the capacity to reveal unexpected relationships within the data. Our findings underscore the potential of LLM-based techniques for supporting qualitative analysis.

% Social media constitutes a rich and influential source of information for qualitative researchers. However, its vast volume and diversity present significant challenges, which can be assisted by computational techniques like topic modelling. But qualitative researchers often struggle to use computational techniques due to a lack of programming expertise and concerns about maintaining the nuanced aspects of their research, such as contextual understanding, subjective interpretations,  and ethical considerations of their data. To address this issue, this thesis explores the integration of BERTopic, an advanced Large Language Model (LLM)-based method, into the Computational Thematic Analysis (CTA) Toolkit to support qualitative data analysis of social media. We conducted interviews and hands-on evaluations in which qualitative researchers compared topics from three modeling techniques --- LDA, NMF and BERTopic. Participants prioritized topic relevance, logical organization, and the capacity to reveal unexpected relationships within the data, valuing detailed, coherent clusters for deeper understanding and actionable insights. BERTopic was favored by 8/12 participants for its ability to uncover hidden connections. These findings underscore the transformative potential of LLM-based tools in providing deeper, more nuanced insights for qualitative analysis of social media data.
\end{abstract}


\ccsdesc[500]{Human Computer Interaction~Generate the Correct Terms for Your Paper}

\ccsdesc{Qualitative Analysis~Generate the Correct Terms for Your Paper}
\ccsdesc[100]{Topic modelling~Generate the Correct Terms for Your Paper}
\ccsdesc[300]{Large Language Model~Generate the Correct Terms for Your Paper}

% \keywords{, , BERT}

\received{12 September 2024}

\maketitle

%======================================================================
\section{Introduction}
%======================================================================

For qualitative researchers, social media serves as a rich and influential source of information \citep{andreotta2019analyzing}. Reddit, in particular, has emerged as a crucial platform for gathering diverse and localized content; offering opinions, advice, and recommendations from trusted sources \citep{ryan2017social, ahn2013social, allen2014social}. This geographically specific and up-to-date information is invaluable for aiding research, synthesis, and presentation of data \citep{dugan2008s, skeels2009social, steinfield2009bowling, morris2010comparison}. Researchers leverage social media to uncover nuanced insights in fields such as mental health, politics, consumer behavior, health and fitness, technology, gaming, education, and social movements \citep{andreotta2019analyzing,ahn2013social, kapoor2018advances}. For instance, public health researchers use Reddit to study health-related topics and understand community perspectives and communication norms, which aids in developing more effective interventions and policies \citep{eysenbach2005patient, heldman2013social, gauthier2022will, rotolo2022hesitancy,gauthier2023agency}.

Despite the availability of social media data, the sheer volume, diversity, noise, dynamic nature, and contextual nuances of this data present significant challenges \citep{andreotta2019analyzing}. Computational techniques like topic modelling are thus necessary to effectively analyze this extensive content \citep{rana2016topic}. However, traditional topic modelling methods like Latent Dirichlet Allocation (LDA),  Biterm Topic Model (BTM), and Non-Negative Matrix Factorization (NMF) often fail to capture nuanced meanings and contextual usage and require extensive data cleaning and pre-processing, making them labor-intensive \citep{yan2013biterm,mazarura2016comparison,zou2016lda}. In contrast, Large Language Model (LLM) based techniques like BERTopic, Top2Vec and GPT-3 offer substantial improvements. They provide advanced understanding of text with minimal pre-processing, adapting efficiently to various contexts \citep{brown2020language, devlin2019bert, vaswani2017attention}.

Further, qualitative researchers face significant challenges when integrating computational techniques like topic modelling into their qualitative workflows  due to a lack of programming expertise \citep{lochmiller2021conducting, baden2022three}. They also have concerns about losing control over the nuanced aspects of their research and doubts about the accuracy of computer-generated insights \citep{abram2020methods,jiang2021supporting,feuston2021putting}. Ethical and privacy concerns further contribute to this reluctance \citep{siiman2023opportunities}. 

Nevertheless, AI tools can significantly aid in analyzing large volumes of social media data. These tools efficiently process and analyze data, identifying trends, sentiments, and patterns that would be challenging to uncover manually \citep{batrinca2015social}. Researchers could utilize AI's capability for deep data analysis, providing a comprehensive understanding of social media interactions \citep{fan2014power}. In particular, topic modelling is increasingly being integrated into qualitative research to manage and interpret large datasets, offering deeper and more objective insights \citep{gillies2022theme}. 

Our research is motivated by the desire to assist qualitative researchers who often face challenges with manual, time-consuming, and potentially biased data analysis methods due to limited programming skills. Recognizing these challenges, we aim to explore the potential of Large Language Models (LLMs) for unsupervised topic modelling, offering a more efficient and objective approach to data analysis. By understanding how qualitative researchers currently use topic modelling and identifying the specific difficulties they encounter, we can develop better tools to support their work. 

We integrated BERTopic into the Computational Thematic Analysis (CTA) toolkit \citep{gauthier2022computational}. Integrating BERTopic into the CTA Toolkit required architectural changes to accommodate the model's advanced requirements which uses word embeddings as compared to the original bag of words methodology, and leverages transformer-based processing. This included incorporating GPU utilization to meet BERTopic’s high computational demands and ensuring efficient processing of large datasets. Additionally, changes to the data filtering methods were necessary to ensure improved data processing and data integrity. 

We then conducted interviews with qualitative researchers to discuss the challenges they face with existing topic modelling tools. Following these discussions, researchers engaged with the CTA toolkit, applying it to their own datasets and models to identify their preferred options. This hands-on exploration allowed them to evaluate the tools in the context of their specific needs. The choices and reasons for their preferences were then collaboratively analyzed, ensuring that the selected topic modelling techniques aligned well with the researchers' needs and preferences. 

Participants prioritized topic relevance, logical organization, and the capacity to reveal unexpected yet significant relationships within the data when evaluating topic modelling techniques. They desired detailed, coherent clusters that clearly separated and grouped significant topics, which facilitated deeper understanding and actionable insights, which were provided by BERTopic. Despite some visualization shortcomings, they valued BERTopic's ability to uncover hidden connections, emphasizing the need for meaningful, comprehensive analysis tools that support their research objectives and enhance data interpretation. BERTopic was ranked first by 8 out of 12 participants (67\%), LDA by 3 out of 12 participants (25\%), and NMF by 1 out of 12 participants (8\%).

In summary, we make three contributions: 

\begin{enumerate}
    \item We integrated BERTopic into the Computational Thematic Analysis Toolkit.

    \item We conducted interviews with qualitative researchers to discuss the challenges they face when using topic modelling tools to analyse social media data.

    \item We discuss the findings on researchers' priorities in topic modelling techniques.
\end{enumerate}

\section{Related Work}
%======================================================================

%\section{HCI Support for Qualitative Research in social media}

The HCI community is increasingly exploring the use of social media to gain insights into human behavior \citep{gauthier2020, gauthier2023agency,andalibi2016understanding,ammari2018pseudonymous}. Topic modelling in particular has been used to support social media analysis by identifying underlying themes and patterns within large datasets, allowing researchers to categorize and summarize vast amounts of unstructured text data efficiently \citep{jacobi2018quantitative,curiskis2020evaluation}. This facilitates the detection of trends, sentiment analysis, and the extraction of meaningful insights from complex and diverse social media content. Topic modelling has been applied across numerous fields -- studies such as those by \citet{jacobi2018quantitative} in journalism, \citet{gauthier2020, gauthier2023agency,rotolo2022hesitancy,eysenbach2005patient,han2021plans} in public health, \citet{haghighi2018using} in urban planning, \citet{bail2018exposure} in political science, and \citet{pousti2021researching} in information systems illustrate the broad applicability and value of topic modelling in leveraging social media data platforms for research.

One such platform is Reddit, which facilitates extensive online interactions through user-generated communities called subreddits, covering diverse topics \citep{kumar2022covid, rocha2023passive}. Its structure, allowing long comments and user anonymity, encourages open and honest discussions \citep{alsinet2021discovering, ammari2018pseudonymous}. Researchers utilize Reddit’s public data for qualitative and quantitative studies, benefiting from its organized subreddit format that simplifies data collection. This makes Reddit a valuable tool for examining online community interactions and behavior, providing rich datasets and insights into various aspects of online communication \citep{proferes2021studying}.

Many studies demonstrate social media's significant potential for mental health support and community engagement \citep{de2014mental, rubya2017video, pretorius2020designing, haimson2015disclosure, wallace2017ageing}. They explore topics like mental health discussions on Reddit, video-mediated peer support, digital interventions, gender transition support, and support for older adults. \citet{gauthier2020} investigates online sobriety communities, while \citet{arif2018acting} and \citet{tsou2015social} highlight social media analytics in crisis management and information operations. These studies collectively underscore social media's diverse impact, emphasizing the importance of advanced topic modelling for deeper insights.

However, the vast and varied nature of social media data presents significant challenges for qualitative researchers \citep{stieglitz2020going}. Many qualitative researchers lack programming or data science expertise, making the adoption of advanced topic modelling tools particularly challenging. While they find machine learning techniques both empowering and exciting, appreciating the novel insights these tools can provide, they also harbor reservations about the trustworthiness of machine-generated results and the potential for missing nuanced content. Studies such as those by \citet{lochmiller2021conducting} and \citet{baden2022three} highlight the dual perspectives of excitement and skepticism researchers feel towards integrating computational methods into their workflows. Additionally, research by \citet{burgess2012twitter}, \citet{lewis2013content}, and \citet{dimaggio2013exploiting} underscores the critical need for intuitive, reliable, and ethically sound computational tools. These tools must address the complexities of social media, bridge the gap for researchers without technical expertise, and respect the methodological foundations of qualitative research.

Traditional topic modelling methods such as LDA \citep{blei2003latent}, Biterm  \citep{yan2013biterm}, NMF \citep{lee1999learning} often fall short in capturing the nuanced meanings and contextual usage of words, necessitating extensive data cleaning and pre-processing, which makes them labor-intensive \citep{xu2015short, hong2010empirical, yan2013biterm,mazarura2016comparison,zou2016lda}. These methods rely on predefined topic numbers and treat words independently, leading to the loss of important contextual information and challenges with short texts like tweets or Reddit comments. Additionally, their effectiveness can be highly sensitive to hyperparameter settings, and they often require substantial computational resources. Conversely, Large Language Model (LLM) based techniques, such as BERTopic \citep{grootendorst2022bertopic}, Top2Vec \citep{angelov2020top2vec} and GPT-3 \citep{brown2020language}, offer significant improvements. These models provide a more advanced understanding of text with minimal pre-processing and adapt efficiently to various contexts \citep{devlin2018bert,brown2020language}. LLMs are leading in topic modelling due to their user-friendliness and advanced capabilities \citep{mu2024large}. Recent research highlights their effectiveness in zero-shot text summarization, achieving near-human performance, suggesting their potential for generating insightful topics \citep{zhang2024benchmarking}.

\section{Topic Modelling Techniques}

%This section examines three prominent topic modelling techniques: Latent Dirichlet Allocation (LDA), Non-negative Matrix Factorization (NMF), and BERTopic. We review the use of these techniques in literature, their specific needs and benefits, and the mathematical foundations behind them. Additionally, we describe how these methods are employed in our quantitative and qualitative analysis, highlighting their effectiveness and identifying current research gaps that merit further exploration.

\subsection{Latent Dirichlet Allocation (LDA)}

Latent Dirichlet Allocation (LDA) is extensively used in the literature due to its robustness and flexibility in identifying hidden thematic structures within large text corpora (e.g., \citet{jelodar2017latent,cohen2014redundancy,zhang2017idoctor,zhao2011comparing}). Since its introduction by \citet{blei2003latent}, LDA has become one of the most popular topic modelling techniques in natural language processing and has been applied in diverse fields such as social media analysis, scientific research, and business intelligence.

LDA has been widely used to analyze Reddit data, demonstrating its effectiveness across various research areas. It has been applied to study sentiment and public opinion on topics like the Russo-Ukrainian conflict and COVID-19 vaccines \citep{guerra2023sentiment, melton2021public}, as well as health discussions on cystic fibrosis, depression, and the pandemic's social impacts \citep{karas2022experiments, tadesse2019detection}. LDA also aids in monitoring public response during crises, such as the COVID-19 pandemic and Jakarta floods \citep{ma2023tracing, rahmadan2020sentiment}, and exploring niche issues like gender-affirming voice interventions \citep{mandava2023characterizing}. Furthermore, it has been used for analyzing suicidal drivers, vaccine sentiment, and other health-related topics on social media \citep{donnelly2023disclosure, melton2021public, khan2023measuring}.

However, LDA's performance is constrained by several inherent limitations. It relies on a ``bag of words" approach, ignoring the order and structure of words in documents, which can dilute the semantic integrity of the topics generated \citep{cordeiro2012twitter}. Furthermore, LDA has been criticized for its inefficiency in handling short texts, common in social media and other contemporary datasets. This inefficiency often leads to less coherent and more general topics, failing to capture the detailed nuances necessary for high-quality insights \citep{bi2018topic}. Literature supports these observations, noting LDA's limitations in topic diversity \citep{egger2022topic}, coherence, and granularity \citep{newman2010automatic, chang2009reading}.

\subsection{Non-Negative Matrix Factorization (NMF)}
Non-negative Matrix Factorization (NMF) is a technique in multivariate analysis and linear algebra where a matrix is factorized into two matrices with non-negative elements, revealing latent structures in the data. Introduced by  \citet{paatero1994positive}, NMF gained significant attention after \citet{lee1999learning} research article, which demonstrated its application in learning parts of objects and face recognition. \citet{lee1999learning} highlight that NMF is particularly effective at clustering data into distinct, easily interpretable groups, supporting its utility for thematic analysis.  

% NMF decomposes a non-negative matrix \( V \) into two non-negative matrices \( W \) and \( H \). The goal of NMF is to find matrices \( W \) and \( H \) such that their product approximates the original matrix \( V \): 

% %The diagram shown in the image represents the NMF process. Here's a breakdown of the components:

% \[
%  V \approx W \times H
% \]

% \begin{itemize}
 
% \item \textbf{\( V \)}: The term-document matrix where rows correspond to terms and columns correspond to documents.
    
% \item \textbf{\( W \)}: The basis matrix which represents the distribution of terms across topics. Each column of \( W \) corresponds to a topic, and each row corresponds to a term.

% \item \textbf{\( H \)}: The coefficient matrix which represents the distribution of topics across documents. Each row of \( H \) corresponds to a topic, and each column corresponds to a document.
 
% \end{itemize}

% The objective of NMF is to minimize the difference between \( V \) and the product \( W \times H \). This can be formulated as an optimization problem:

% \[
% \min_{W, H} \| V - W H \|_F
% \]

% where \( \| \cdot \|_F \) denotes the Frobenius norm, which measures the element-wise difference between the matrices.

% By decomposing \( V \) into \( W \) and \( H \), NMF provides a low-rank approximation that captures the underlying structure of the data. This makes it a powerful tool for applications such as topic modelling, image processing, and collaborative filtering. This method is particularly useful for uncovering hidden patterns in data \citep{lee1999learning, lee2001algorithms}.

For topic modelling and short-text analysis in Reddit, NMF effectively decomposes text data into meaningful topics using term-document matrices \citep{egger2022topic, albalawi2020using}. It handles emotion and sentiment analysis well, comparing favorably with traditional LDA methods, and is used to detect and analyze communities by integrating sentiment analysis \citep{curiskis2020evaluation}. In semantic and social analysis, NMF models user participation and content data effectively and extends to analyze multiplex networks, integrating information from multiple Reddit communities \citep{wu2008social}. NMF is also used in climate change discussions on Reddit, effectively handling the diverse and dynamic nature of social media posts to extract meaningful insights \citep{parsa2022analyzing}. Additionally, NMF has been applied to mental health analysis, identifying latent structures in text data to understand issues like depression, stress, and suicide risk on social media platforms \citep{garg2023mental}.

\subsection{BERTopic}

BERTopic, introduced by Maarten Grootendorst \citep{grootendorst2022bertopic}, is a state-of-the-art topic modelling technique that leverages transformers and class-based term frequency-inverse document frequency (c-TF-IDF) to create dense embeddings and hierarchical topic representations. This integration allows for more nuanced and contextually aware topic modelling, which is particularly beneficial for analyzing complex and large-scale datasets \citep{reimers2019sentence}. This method benefits from BERT's advanced capabilities in capturing contextual information through deep bidirectional transformers, providing more accurate and nuanced topic modelling compared to traditional models like LDA and NMF \citep{devlin2019bert}. The foundational advancements of the Transformer model \citep{vaswani2017attention} further enhance BERTopic's effectiveness in providing refined and contextually aware topic modelling. 

 The implementation leverages advanced techniques in natural language processing and machine learning, including the Sentence-BERT model, UMAP for dimensionality reduction, HDBSCAN for clustering, and cTF-IDF for topic extraction. The token sets are prepared by converting them into a text format suitable for the BERTopic model, and then generating dense vector embeddings.

\section{Summary}

LDA and NMF are widely used for topic modelling in social media data analysis, but they have several limitations. LDA assumes a known number of topics and that words are generated independently given the topic, which can lead to loss of context and struggles with short texts like tweets or Reddit comments \citep{blei2003latent,o2015down}. Additionally, LDA’s performance is sensitive to hyperparameters and requires substantial computational resources \citep{wallach2009evaluation}. NMF, while useful for producing sparse, interpretable results, is sensitive to matrix initialization and requires careful tuning, and it does not inherently capture probabilistic topic distributions \citep{lee1999learning, lee2001algorithms}. Studies show mixed results for these methods on social media data, highlighting the need for better algorithms to gain more accurate and insightful analyses \citep{o2015down}.

Comparative studies, such as those by \citet{egger2022topic} demonstrated that BERTopic outperforms LDA, NMF, and Top2Vec in capturing semantic nuances and contextual information particularly on Twitter. Consequently, we incorporated BERTopic into our study to validate its effectiveness in enhancing our ability to extract nuanced insights and understand the complexities of social media data, specifically on Reddit. This approach is particularly beneficial for qualitative researchers from a Human-Computer Interaction (HCI) perspective, as it addresses the need for more intuitive and contextually rich analysis tools.

\section{The Computational Thematic Analysis Toolkit}
%======================================================================

The Computational Thematic Analysis (CTA) Toolkit, developed by \citet{gauthier2022computational}, integrates qualitative thematic analysis and computational methods. We chose the CTA Toolkit as our experimental platform because its functionality aligned well with our research requirements; it provides a visual interface for qualitative analysis of social media data using computational techniques like topic modelling. We extended the toolkit by adding BERTopic as a topic modelling option, which allowed us to directly compare it to LDA and NMF under a single interface. We now provide a brief overview of the CTA toolkit's key features for data collection, data cleaning and filtering, and modelling before describing how we integrated BERTopic.

%\section{Data Collection}
The data collection module automates the retrieval and organization of data from online platforms like Twitter and Reddit. Researchers import data by defining parameters such as time periods and specific communities. The toolkit groups submissions and their corresponding comments into cohesive discussions, ensuring that the data is organized and ready for subsequent analysis. To address the ethical complexities inherent in online data collection, the toolkit prompts researchers to consider the ethical implications of their work before data is imported for analysis. These prompts guide researchers through considerations such as community consent and the handling of deleted posts. 

The data cleaning and filtering module then employs natural language processing (NLP) tools, including NLTK and spaCy, to tokenize, stem, and lemmatize the text as it is imported. It also tags parts of speech and identifies stop words, facilitating a detailed and precise cleaning process. The toolkit enhances transparency by displaying a summary of the tokens and filtering rules applied, allowing researchers to see the impact of these steps on their data. This feature is essential for maintaining the integrity and reproducibility of the research.

%\section{Modeling \& Sampling}

\begin{figure}[tb!]
    \centering
    \includegraphics[width=1\textwidth]{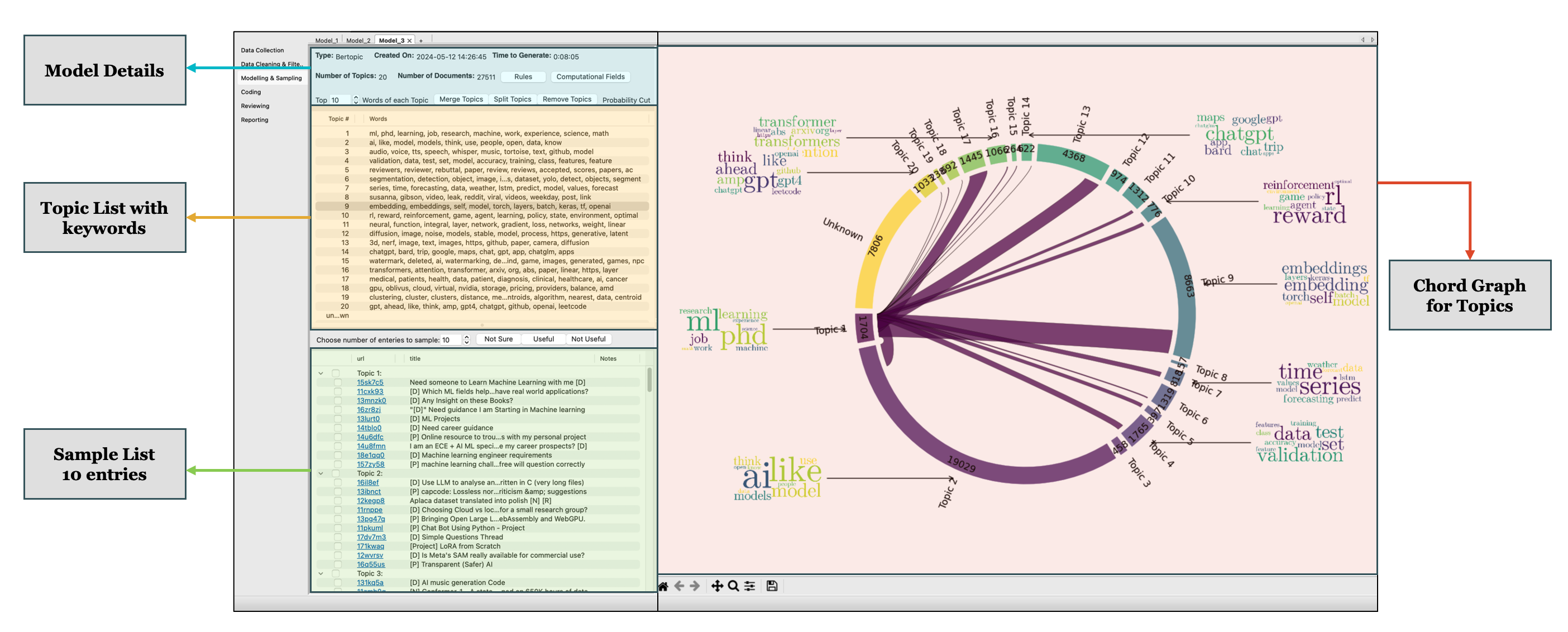}
    \caption[Modeling and Sampling Module of CTA with BERTopic Integration]{ Modeling and Sampling Module in the CTA Toolkit with BERTopic Integration: Model Details, a Topic List with keywords, Sample List of entries on the left side, and a Chord Graph for with Topics as Word clouds on the right. This layout facilitates detailed and comprehensive analysis of topic modeling results. }
    \label{fig:bertopic_window}
\end{figure}

The modeling and sampling module (\autoref{fig:bertopic_window}) allows researchers to interactively select and review data models, ensuring that the most relevant and representative data is used in the thematic analysis. This interactive approach supports the iterative nature of qualitative research, allowing researchers to refine their models based on emerging insights. To identify latent patterns within large datasets the CTA toolkit uses unsupervised topic modeling techniques --- LDA \citep{blei2003latent}, biterm \citep{yan2013biterm}, NMF \citep{lee1999learning}. The toolkit's visual interface includes a chord diagram visualization of generated topic models. The outer ring represents the different topics identified in the dataset. Each segment is labeled with a topic number (e.g., Topic 1, Topic 2, etc.). Alongside each topic label, there are word clouds showing the most relevant terms for each topic. The inner lines connecting different segments represent the number of overlapping documents between topics. Thicker lines imply more shared documents between the connected topics. 

%Interpreting the visualizations from \autoref{fig:explain}: Each topic is represented by a segment on the ring with a word cloud indicating the most significant words within that topic. For example, Topic 7 might include terms like ``leather," ``vegan," ``plant," suggesting this topic revolves around sustainable and vegan leather products. The chords (lines) between topics show how they are related. For example, if there is a thick chord between Topic 14 and Topic 32, it means these topics share many common documents or are closely related in their themes (hemp, cotton, bamboo are sustainable materials). 

%Specific examples from the diagram include Topic 19, where words such as ``circular," ``economy," ``kindly" imply a focus on circular economy practices in sustainable fashion.

% \begin{figure}[tb!]
%     \centering
%     \includegraphics[width=1\linewidth]{P6/bertopic_Sfashion.png}
%     \caption{Topic keywords for BERTopic on the r/SustainableFashion dataset}
%     \label{fig:explain}
% \end{figure}

% \begin{figure}[!h]
%     \centering
%     \includegraphics[width=1\textwidth]{images/step5.png}
%     \caption{The addition of BERTopic, a transformers-based hierarchical topic modeling method for efficient exploratory analysis.}
%     \label{fig:bertopic_integrate}
% \end{figure}

%\subsection{Integration of BERTopic}

We integrated BERTopic within the CTA Toolkit's pipeline. This work posed several challenges, primarily due to the inherent differences between BERTopic's LLM-based processing requirements and the existing bag-of-words architecture. Additionally, BERTopic's reliance on transformer models demands higher computational resources, including increased memory and processing power. For BERTopic, we leveraged the device's built-in GPU to enhance processing efficiency. Moreover, the toolkit included filters and preprocessing steps tailored for bag-of-words models, which were not suitable for BERTopic's needs. The preprocessing step involved default filters which eliminate terms with TF-IDF values less than 75\%. This approach aimed to retain only the most significant terms; however, it led to the loss of almost 80\% of the data, which proved to be detrimental to the analysis. Therefore, we decided to eliminate this filtering step for BERTopic to preserve the integrity and comprehensiveness of the data.

\section{Interviews with Qualitative Researchers}

%\section{Introduction}

To understand the specific needs and challenges faced by qualitative researchers, we conducted a series of interviews with experts. We conducted interviews to identify researchers' requirements and gather insights into their experiences with traditional topic modelling techniques, including manual coding methods, or using proprietary software, and their impressions of modern unsupervised topic modelling techniques like BERTopic.

We interviewed 12 participants, each an expert in qualitative research with distinct interests and varying levels of experience in their respective fields.  We asked each participant to identify a subreddit of interest, downloaded the archived data, and loaded it into the Computational Thematic Analysis Toolkit. The datasets spanned a wide range of domains, including gaming, technology, COVID-19, health, environment, and politics. 

We then asked participants to reflect on their data set through three unsupervised topic models:  LDA, NMF, and BERTopic. The three topic modelling techniques were selected to reflect different conceptual approaches. LDA is a probabilistic approach which is widely used in the literature (e.g. \cite{Grimmer2010PublicOpinion, Hoffman2010OnlineLDA, Teh2006HDP}). NMF offers computational efficiency and a matrix-based perspective that is fast and effective for large datasets \citep{wu2008social}. BERTopic leverages cutting-edge capabilities of language models to provide deep, contextually rich topic modelling \citep{grootendorst2022bertopic}. We asked participants to examine their data to help explain how the topic models might be useful. 

This study has undergone a thorough review process and received ethics clearance from our Research Ethics Board (REB \#46062) from University. The review process involved a detailed examination of the study's methodology, participant recruitment procedures, data collection methods, and potential risks to ensure compliance with ethical standards.

% The main study involved conducting twelve interviews with qualitative researchers to identify their needs and challenges in topic modelling. The interviews evaluated their experiences with traditional methods and introduced an automated, interactive interface that utilizes both,  LLM based technique - BERTopic and traditional techniques LDA and NMF for real-time, unsupervised topic modelling. 

\subsubsection{Participants}

Participants were selected through purposive sampling to ensure that they had relevant experience and expertise in the fields under investigation. The selection criteria included a minimum level of experience in qualitative research and active involvement conducting content analysis in their work. The demographic composition of the participants included a balanced representation of gender, age, and occupational backgrounds. This diversity ensured that the study captured a wide range of perspectives and insights, enhancing the richness and depth of the analysis.

Participants were recruited through direct invitations sent via email, leveraging professional networks and academic contacts. This recruitment method ensured that the participants were highly motivated and well-qualified to contribute to the study. Their diverse backgrounds and research interests provided a comprehensive understanding of the various topics discussed on the selected subreddits. \autoref{tab:participant_details} summarizes participant information, detailing the subreddit they analyzed, their research area and years of experience.

\begin{table}[tb!]
    \centering
    \caption[Participant Details and Research Areas]{An overview of the Reddit dataset used, participant IDs, their respective research areas, and years of experience.}
    \label{tab:participant_details}
    \begin{tabular}{p{4cm} p{1cm} p{6cm} p{3cm} }
      \toprule
        \textbf{Reddit Dataset} & \textbf{PID} & \textbf{Research Area} & \textbf{Years Experience} \\ 
        \midrule        
        r/disabledgamers & P1 & Accessibility/disability research & 10 years \\ 
        r/MachineLearning & P2 & eXplainable AI in LLMs & 3 years \\ 
        r/HermanCainAward & P3 & Sentimental analysis on social media & 3 years \\ 
        r/cybersecurity & P4 & Cyber security and privacy risk in social robot & 6 months \\ 
        r/emergencymedicine & P5 & Health policy and emergency care & 12 years \\ 
        r/SustainableFashion & P6 & Fashion and sustainability management & 10 years \\ 
        r/LanguageTechnology & P7 & Large Language Models for GenAI & 8 years \\ 
        r/nutrition & P8 & Nutrition and Diet quality & 2 years \\ 
        r/uwaterloo & P9 & Machine Learning & 4 years \\ 
        r/Coronavirus & P10 & Covid 19 vaccine hesitancy & 2 years \\ 
        r/bipolar & P11 & Care quality for bipolar disorder & 6 years \\ 
        r/CoronavirusCanada & P12 & STEM and medicine & 5 years \\ 
        \bottomrule
    \end{tabular}
\end{table}

\subsubsection{Procedure}

We directly emailed researchers to explain the study's aims and invite their participation. Upon indicating their interest in participation, we identified an appropriate subreddit for analysis based on their research interests, obtained signed consent letters, and scheduled a mutually convenient date and time for the interview, which could be conducted either online or in-person meeting.

On the day of the study, we initiated the session with a presentation of approximately five minutes. This presentation provided a comprehensive overview of the project, outlining its objectives and the significance of the research. We also addressed any questions or clarifications raised by the participants, ensuring they had a thorough understanding of the study's context. 

%\autoref{fig:interview} describes the interview procedure we followed. 

% \begin{figure}
%     \centering
%     \includegraphics[width=1\textwidth]{images/interview.png}
%     \caption[Interview phases: Backgrounds, Model Comparison, Discussion]{Background Interview phase involved conducting interviews using questionnaires based on users’ research backgrounds. The demonstration phase involved comparing and contrasting the results of different models (LDA, NMF, BERTopic) on the Computational Thematic Analysis Toolkit. Lastly, the interview on topic models phase analyzed discussions with qualitative researchers on their criteria for ranking the models and their overall experience with the three approaches.}
%     \label{fig:interview}
% \end{figure}

\subsubsection{Background Interview}
In the background interview phase, we aimed to gather comprehensive information about the participants' research background and their application of content analysis techniques. The interview commenced with questions about their main area of research and expertise, as well as the duration of their involvement in their respective fields. We then explored the relevance of content analysis in their research by inquiring where and how they utilize these techniques, supplemented by requests for specific use cases. The discussion progressed to understanding the types of data typically used for content analysis, including the nature and sources of these data—whether public, private, or institutional. Participants were asked to detail the methods and tools they currently employ for content analysis, including any specific software or libraries. This was followed by a focus on the challenges they encounter with existing tools and their suggestions for improvements. To conclude this phase, we sought their insights on the potential benefits of integrating modern technologies such as AI, Large Language Models (LLMs), and Machine Learning into their research practices. 

%\autoref{tab:interview_phase1} presents the list of interview questions we asked the participants in background interview phase.

Following this preliminary discussion, we proceeded to demonstrate the CTA toolkit. Initially, we presented statistical information pertinent to the selected subreddit and outlined the specific tasks the participants would undertake. We then demonstrated the key features of the toolkit, describing how it integrates various topic modelling techniques, including BERTopic, LDA, and NMF. Participants were then provided access to the software, with a 10-20 minute period allocated for them to explore its functionalities. During this time, they examined the three models in depth, comparing the topics generated by each.

\subsubsection{Interview on Topic Models}

In Interview on Topic Models phase, conducted after participants had experimented with the CTA toolkit, we focused on gathering their detailed feedback on its efficacy and relevance to their work. Initially, we asked participants to use the Microsoft Desirability Toolkit to select words that best described our method, with specific focus areas for the overall visualization experience, LDA, NMF, and BERTopic models. Participants were then requested to rank the three models (LDA, NMF, BERTopic) based on their usefulness, explaining the factors influencing their rankings. We sought comparisons between our proposed method and existing tools that the participants had previously used, asking them to highlight specific features they liked or disliked and to suggest areas for improvement.

Additionally, we investigated the relevance of the three models to their research, asking which model suited their work the best and why. We explored the potential integration of our methodology into their work, including any anticipated challenges. Finally, participants were asked to rate the likelihood of incorporating our model into their working pipeline on a scale of 1-10 and to share any additional insights about the proposed methodology and the interview process. This phase aimed to comprehensively understand the practical utility and potential improvements for our approach based on direct user feedback.

\subsection{Equipment}

Model training was conducted on a MacBook Air with an Apple M1 chip, featuring an 8-core CPU (4 performance and 4 efficiency cores), a 7-core GPU, and 16GB of RAM for efficient processing of large datasets and complex computations. The setup also included a Philip's 4K Ultra HD LCD monitor (28" inch) to facilitate detailed visualization of results and model performance.

\subsection{Participant Data Sets} \label{data_collect}

We gathered data from 12 subreddits covering six broad topics: gaming, technology, COVID-19, health, and the environment (\autoref{tab:dataset_descriptions}). The data from PushShift.io \citep{baumgartner2020pushshift} was decompressed from the \texttt{.zst} (Zstandard) file format to \texttt{.json} format and organized into a structured month-year format for ease of access and analysis.  Processing times varied based on dataset size, ranging from approximately 10 minutes to 2 hours.

During these discussions, we consulted with the participants to determine if they had specific dates or years in mind for analysis. If they did, we collected data for that specified duration. Otherwise, we utilized the entire dataset available on the subreddit.

\begin{sidewaystable}
    %\scriptsize % Set font size for the table
    \renewcommand{\arraystretch}{1.15} % Increase row spacing
    %\centering
    \vspace*{15cm}
    \begin{tabular}{p{2.8cm} p{0.6cm} p{1.75cm} p{2.0cm} p{2.0cm} p{6.3cm} p{1.2cm}}
        \toprule
        \textbf{Subreddit} & \textbf{ID} & \textbf{Number of Documents} & \textbf{Start Date} & \textbf{End Date} & \textbf{Description} & \textbf{Rank by Size * } \\ 
        \midrule
        
        r/disabledgamers & P1 & 4,403 & 01/01/2015 & 31/12/2023 & A community for gamers with disabilities to discuss accessibility options, adaptive controls, and gaming experiences. & Top 7\% \\
        
        r/MachineLearning & P2 & 27,506 & 01/01/2023 & 31/12/2023 & A vibrant community dedicated to discussions on machine learning, AI, and data science. & Top 1\% \\ 

        r/HermanCainAward & P3 & 35,013 & 01/09/2021 & 31/12/2023 & A subreddit documenting and discussing cases of COVID-19 skepticism and its consequences. & Top 1\% \\ 

        r/cybersecurity & P4 & 26,692 & 01/01/2023 & 31/12/2023 & A resource for cybersecurity topics including threat analysis, security breaches, and industry best practices. & Top 1\% \\ 
    
        r/emergencymedicine & P5 & 14,117 & 01/04/2012 & 31/12/2023 & A community focused on discussions about emergency medical care, including case studies and clinical guidelines. & Top 10\% \\ 
    
        r/SustainableFashion & P6 & 5,845 & 01/05/2017 & 31/12/2023 & A platform for discussing sustainable fashion practices, upcycling, and eco-friendly brands. & Top 3\% \\ 
    
        r/LanguageTechnology & P7 & 13,420 & 01/10/2011 & 31/12/2023 & A subreddit focusing on advancements in NLP, computational linguistics, and language-based AI. & Top 3\% \\ 
    
        r/nutrition & P8 & 14,313 & 01/01/2023 & 31/12/2023 & A community discussing various topics related to nutrition, diet plans, and health outcomes. & Top 1\% \\ 
    
        r/uwaterloo & P9 & 15,793 & 01/01/2023 & 31/12/2023 & A community for discussions related to academic life, programs, and events at the University of Waterloo. & Top 10\% \\ 
    
        r/Coronavirus & P10 & 23,906 & 01/11/2020 & 01/02/2021 & A large community dedicated to global news, scientific research, and public health guidelines related to COVID-19. & Top 1\% \\ 
   
        r/bipolar & P11 & 24,512 & 01/06/2023 & 31/12/2023 & A supportive community for individuals affected by bipolar disorder, offering personal experiences and coping strategies. & Top 5\% \\ 
    
        r/CoronavirusCanada & P12 & 26,862 & 01/03/2020 & 31/12/2023 & A subreddit focusing on COVID-19 discussions within Canada, including local news and government policies. & Top 5\% \\ 
        \bottomrule
    \end{tabular}
    \caption[Dataset Descriptions for Selected Subreddits]{Summaries of data collected from official Reddit pages of various subreddits.}
    \label{tab:dataset_descriptions}
\end{sidewaystable}

\section{Topic Modelling}

We used a standard procedure to create models for each subreddit, following the same protocol for LDA, NMF, and BERTopic. This process involved performing tokenization, which were then converted into formats suitable for each model: a bag-of-words corpus for LDA, a TF-IDF matrix for NMF, and dense vector embeddings for BERTopic. Each model was configured according to its specific requirements, with hyperparameters and settings tailored to optimize performance, and kept to default settings, as detailed in the following sections.

We began by merging submissions and comments from JSON files (\texttt{RS\_<filename>.json} and \texttt{RC\_<filename>.json}) in a specified subreddit directory. Submissions and comments are merged into threads using their IDs, creating a cohesive narrative for each discussion. Threads are saved as JSON for analysis. We then tokenized each thread with \texttt{gensim}'s \texttt{simple\_preprocess}, and stored tokens in a dictionary. Stopwords and irrelevant terms are removed using NLTK, while bigrams are identified with \texttt{gensim}.

\textbf{LDA Implementation}: The LDA model was implemented using the \texttt{gensim} library. For tokenization and preprocessing, \texttt{gensim} handled tokenization and bigrams, \texttt{NLTK} removed stopwords, and \texttt{spaCy} performed lemmatization. A dictionary mapped unique words to IDs, and token sets were converted to a bag-of-words corpus. Hyperparameters \(\alpha\) (symmetric) and \(\eta\) (auto) were chosen to balance topic distribution.
  
\textbf{NMF Implementation}: The NMF model was implemented using the \texttt{scikit-learn} library. Token sets were converted into space-separated strings for the TF-IDF vectorizer. The vectorizer transformed text into a TF-IDF matrix, which was saved and fitted to the NMF model, with a set random state for reproducibility.

\textbf{BERTopic Implementation}: BERTopic used the \texttt{sentence-transformers} library for embeddings, \texttt{umap-learn} for dimensionality reduction, and \texttt{hdbscan} for clustering. The \texttt{bertopic} library integrated these components, facilitating effective topic modelling analysis.

%Lemmatization, acronym expansion, and cleaning steps ensure data readiness for topic modelling.

The number of topics (\texttt{num\_topics}) for LDA and NMF were chosen based on the maximum topic coherence value observed from generating topic models ranging from 5 to 50 topics, in increments of 5. To determine an optimal value BERTopic, we generated 11 topic models and chose the median value. In doing so, we observed that the mean and median values for \texttt{nr\_topics} were almost in conjunction, differing by average standard deviation between the median and average values being $2.09 \pm 1.37$ across 12 datasets.

\subsection{Data Collection and Analysis}

During our study, we offered participants the option to conduct interviews through video chat or in person at the university campus. All participants chose the online format, conducted through Microsoft Teams, a trusted platform with end-to-end encryption. Each interview lasted between 1 - 1.5 hours. With participants' consent, we recorded the audio for transcription using the platform's built-in feature, and recordings were deleted within 24 hours to ensure privacy. The transcriptions included responses to the phase 1 and 2 questionnaires. For analysis, we systematically reviewed the transcriptions, identifying themes and key aspects that reflected participants' interactions with the topic models, thereby capturing the nuances of their experiences.

In our analysis comparing the ranking given by the participants for the three topic modelling methods, we utilized the \textbf{Friedman test} and the \textbf{Nemenyi post-hoc test} for statistical inference. The Friedman test was suitable for comparing the ranks of multiple models on the same datasets, revealing overall significant differences without assuming normal distribution. The Nemenyi post-hoc test complemented this by identifying specific pairs of models with statistically significant differences, adjusting for multiple comparisons. 

Additionally, we employed the \textbf{ANOVA (Analysis of Variance)} and \textbf{Tukey's Honest Significant Difference (HSD)} tests for statistical inference. ANOVA was chosen because it is ideal for comparing the means of multiple groups and determining if statistically significant differences exist among the models. This flexibility makes ANOVA applicable to various metrics, including Number of Topics, Topic Coherence, Topic Diversity, KL-Divergence, Perplexity, and Execution Time. ANOVA combined with post-hoc test Tukey's HSD provides a consistent and thorough method for comparative performance analysis, determining which specific group means are different. It controls the Type I error rate and provides confidence intervals for the differences between means.

\section{Results}

BERTopic consistently outperformed LDA and NMF in terms of quantitative metrics, including topic coherence, topic diversity, and KL divergence. On the other hand, and as expected, NMF was the least computationally-demanding model and required the lowest amount of time to compute. LDA tended to lay in-between the other two models across measures. Participant preferences closely align with these findings. BERTopic was ranked first by 8 out of 12 participants (67\%), LDA by 3 out of 12 participants (25\%), and NMF by 1 out of 12 participants (8\%). A Friedman test revealed that these differences were statistically significant, \(\chi^2_{(2)} = 108\), \(p < 0.001\). Post-hoc analysis using the Nemenyi test showed significant differences between BERTopic and LDA (\(p < 0.05\)), BERTopic and NMF (\(p < 0.01\)), but not between LDA and NMF (\(p > 0.05\)).

We first present these quantitative results in detail. Then, in presenting our analysis of interviews with researchers, we highlight rationale for these preferences alongside practical concerns for each topic modelling technique.

%The test statistic \(Q = 108\) with \(df = 2\) follows a chi-square distribution, leading to a statistically significant result with \(p < 0.001\), highlighting the variance in participants' model preferences.

%(Most favoured to Least favoured): BERTopic \(\{8, 1, 3\}\), LDA \(\{3, 3, 6\}\), and NMF \(\{1, 8, 3\}\).

%\section{Quantitaive results}
%This section of the chapter focuses on quantitative data analysis using various metrics such as topic coherence, diversity, KL divergence, and execution time. Alongside these metrics, the results from the Microsoft Desirability Toolkit are analyzed to provide a comprehensive evaluation of the models.

\subsection{Number of Topics}

\begin{table}[tb!]
    \centering
    \caption[Summary of Results: Number of topics]{Number of topics and descriptive statistics in the Reddit datasets using LDA, NMF, and BERTopic}
    \label{tab:reddit_topics}
    \begin{tabular}{ p{5cm} p{3cm} p{3cm} p{3cm} }
        \toprule
        \textbf{Reddit Dataset } & \textbf{LDA} & \textbf{NMF} & \textbf{BERTopic} \\ 
        \midrule
        r/disabledgamers & 10 & 5 & 22 \\
        r/MachineLearning & 10 & 5 & 56 \\
        r/HermanCainAward & 40 & 40 & 131 \\
        r/cybersecurity & 35 & 5 & 142 \\
        r/emergencymedicine & 30 & 50 & 20 \\
        r/SustainableFashion & 15 & 10 & 35 \\
        r/LanguageTechnology & 30 & 20 & 78 \\
        r/nutrition & 10 & 5 & 27 \\
        r/uwaterloo & 15 & 10 & 43 \\
        r/Coronavirus & 10 & 10 & 133 \\
        r/bipolar & 45 & 10 & 56 \\
        r/CoronavirusCanada & 10 & 15 & 62 \\
        \midrule
        \textbf{Mean} & 22 & 15 & 67 \\
        \textbf{Minimum} & 10 & 5 & 20 \\
        \textbf{Maximum} & 45 & 50 & 142 \\
        \bottomrule
    \end{tabular}
\end{table}

BERTopic consistently identified the largest number of distinct topics across the datasets, with a mean of \texttt{67} topics, ranging from \texttt{20} to \texttt{142}. LDA follows with a mean of \texttt{22} topics, with a minimum of \texttt{10} and a maximum of \texttt{45}. NMF identifies the fewest topics on average, with a mean of \texttt{15}, a minimum of \texttt{5}, and a maximum of \texttt{50}.
 \autoref{tab:reddit_topics} compares the topic coherence scores of these three topic modelling methods.

A one-way ANOVA revealed a significant effect of the topic modelling method on the number of topics identified, $F_{2, 33} = 12.00$, $p = .00012$, $\eta^2 = .42$. To explore these differences further, a Tukey HSD post hoc test was conducted. The Tukey HSD test indicated that the mean number of topics for LDA ($M = 22$) was significantly lower than BERTopic ($M = 67.00$), with a mean difference of 45.41 (95\% CI, 17.16 to 73.66), $p = .001$. Additionally, NMF ($M = 15$) was also significantly lower than BERTopic ($M = 67.00$), with a mean difference of 51.66 (95\% CI, 23.41 to 79.9), $p = .0002$. However, NMF did not significantly differ from LDA ($p = 0.85$). 

%In summary, BERTopic demonstrated a significantly higher number of topics compared to both LDA and NMF, while no significant differences were observed between LDA and NMF.

\subsection{Topic Coherence}
\begin{table}[tb!]
    \centering
    \caption[Summary of Results: Topic Coherence]{Topic Coherence in the Reddit datasets using LDA, NMF, and BERTopic}
    \label{tab:reddit_coherence}
    \begin{tabular}{ p{5cm} p{3cm} p{3cm} p{3cm} }
        \toprule
        \textbf{Reddit Dataset} & \textbf{LDA} & \textbf{NMF} & \textbf{BERTopic} \\ 
        \midrule
        r/disabledgamers       & 0.494 & 0.667 & 0.639 \\
        r/MachineLearning      & 0.475 & 0.721 & 0.649 \\
        r/HermanCainAward      & 0.456 & 0.544 & 0.575 \\
        r/cybersecurity        & 0.520 & 0.714 & 0.687 \\
        r/emergencymedicine    & 0.460 & 0.527 & 0.623 \\
        r/SustainableFashion   & 0.386 & 0.573 & 0.685 \\
        r/LanguageTechnology   & 0.474 & 0.599 & 0.596 \\
        r/nutrition            & 0.543 & 0.849 & 0.683 \\
        r/uwaterloo            & 0.507 & 0.706 & 0.695 \\
        r/Coronavirus          & 0.564 & 0.748 & 0.683 \\
        r/bipolar              & 0.543 & 0.819 & 0.573 \\
        r/CoronavirusCanada    & 0.584 & 0.746 & 0.670 \\
        \midrule
        \textbf{Mean}          & 0.500 & 0.684 & 0.647 \\
        \textbf{Minimum}       & 0.386 & 0.527 & 0.573 \\
        \textbf{Maximum}       & 0.584 & 0.849 & 0.695 \\
        \bottomrule
    \end{tabular}
\end{table}

 LDA exhibits coherence scores ranging from a minimum of \texttt{0.386} to a maximum of \texttt{0.584}, with a mean score of \texttt{0.500}. NMF shows a broader range of coherence scores, from \texttt{0.527} to \texttt{0.849}, with a higher mean score of \texttt{0.684}. BERTopic, meanwhile, has a relatively consistent range of coherence scores, between \texttt{0.573} and \texttt{0.695}, with a mean score of \texttt{0.647}.
 \autoref{tab:reddit_coherence} compares the topic coherence scores of these three topic modelling methods.

A one-way ANOVA revealed a significant effect of the topic modelling method on topic coherence, $F_{2, 33} = 21.25$, $p < .001$, $\eta^2 = .563$. To explore these differences further, a Tukey HSD post hoc test was conducted. The Tukey HSD test indicated that the mean topic coherence for LDA ($M = 0.500$) was significantly lower than BERTopic ($M = 0.647$), with a mean difference of 0.147 (95\% CI, 0.038 to 0.256), $p = .011$. Additionally, LDA was also significantly lower than NMF ($M = 0.684$), with a mean difference of 0.184 (95\% CI, 0.076 to 0.293), $p = .002$. However, NMF did not significantly differ from BERTopic, with a mean difference of -0.037 (95\% CI, -0.146 to 0.071), $p = .719$. 

%In summary, both NMF and BERTopic demonstrated significantly higher topic coherence compared to LDA, while no significant differences were observed between NMF and BERTopic.

\subsection{Topic Diversity}

\begin{table}[tb!]
    \centering
    \caption[Summary of Results: Topic Diversity]{Topic Diversity in the Reddit datasets using LDA, NMF, and BERTopic, higher value is better}
    \label{tab:reddit_diversity}
    \begin{tabular}{p{5cm} p{3cm} p{3cm} p{3cm}}
        \toprule
        \textbf{Reddit Dataset} & \textbf{LDA} & \textbf{NMF} & \textbf{BERTopic} \\ 
        \midrule
        r/disabledgamers       & 0.680 & 0.920 & 0.967 \\
        r/MachineLearning      & 0.770 & 0.880 & 1.000 \\
        r/HermanCainAward      & 0.723 & 0.828 & 0.990 \\
        r/cybersecurity        & 0.751 & 0.900 & 1.000 \\
        r/emergencymedicine    & 0.673 & 0.796 & 1.000 \\
        r/SustainableFashion   & 0.653 & 0.880 & 1.000 \\
        r/LanguageTechnology   & 0.700 & 0.900 & 1.000 \\
        r/nutrition            & 0.800 & 0.900 & 1.000 \\
        r/uwaterloo            & 0.773 & 0.870 & 1.000 \\
        r/Coronavirus          & 0.750 & 0.920 & 1.000 \\
        r/bipolar              & 0.727 & 0.920 & 0.992 \\
        r/CoronavirusCanada    & 0.800 & 0.920 & 1.000 \\
        \midrule
        \textbf{Mean}          & 0.733 & 0.866 & 0.995 \\
        \textbf{Minimum}       & 0.653 & 0.796 & 0.967 \\
        \textbf{Maximum}       & 0.800 & 0.920 & 1.000 \\
        \bottomrule
    \end{tabular}
\end{table}
 LDA exhibits diversity scores ranging from a minimum of \texttt{0.653} to a maximum of \texttt{0.800}, with a mean score of \texttt{0.733}. NMF shows a range of diversity scores from \texttt{0.796} to \texttt{0.920}, with a mean score of \texttt{0.866}. BERTopic, meanwhile, has a high and consistent range of diversity scores, between \texttt{0.967} and \texttt{1.000}, with a mean score of \texttt{0.995}. \autoref{tab:reddit_diversity} compares the topic diversity scores of these three topic modelling methods.

%The data indicate that BERTopic achieves the highest topic diversity scores, suggesting it can capture a wide variety of subtopics within the datasets. NMF also shows high diversity scores, albeit with more variability compared to BERTopic. LDA has the lowest diversity scores among the three methods, indicating a more constrained range of subtopics. 

A one-way ANOVA revealed a significant effect of the topic modelling method on topic diversity, $F_{2, 33} = 154.13$, $p < .001$, $\eta^2 = .903$. To explore these differences further, a Tukey HSD post hoc test was conducted. The Tukey HSD test indicated that the mean topic diversity for LDA ($M = 0.733$) was significantly lower than BERTopic ($M = 0.995$), with a mean difference of 0.262 (95\% CI, 0.226 to 0.299), $p < .001$. Additionally, NMF ($M = 0.866$) was also significantly lower than BERTopic, with a mean difference of 0.1096 (95\% CI, 0.072 to 0.146), $p < .001$. NMF also significantly differed from LDA, with a mean difference of 0.152 (95\% CI, 0.115 to 0.189), $p < .001$. 

%In summary, BERTopic demonstrated a significantly higher topic diversity compared to both LDA and NMF, while NMF also demonstrated significantly higher topic diversity compared to LDA.

\subsection{KL Divergence}

\begin{table}[tb!]
    \centering
    \caption[Summary of Results: KL-Divergence]{KL-Divergence in the Reddit datasets using LDA, NMF, and BERTopic, lower value is better}
    \label{tab:reddit_kl_divergence}
    \begin{tabular}{ p{5cm} p{3cm} p{3cm} p{3cm} }
        \toprule
        \textbf{Reddit Dataset } & \textbf{LDA} & \textbf{NMF} & \textbf{BERTopic} \\ 
        \midrule
        r/disabledgamers       & 0.020 & 7.956 & 0.004 \\
        r/MachineLearning      & 0.095 & 8.397 & 0.003 \\
        r/HermanCainAward      & 0.109 & 10.166 & 0.027 \\
        r/cybersecurity        & 0.093 & 9.479 & 0.002 \\
        r/emergencymedicine    & 0.048 & 11.766 & 0.014 \\
        r/SustainableFashion   & 0.125 & 11.042 & 0.012 \\
        r/LanguageTechnology   & 0.144 & 9.607 & 0.001 \\
        r/nutrition            & 0.027 & 7.298 & 0.003 \\
        r/uwaterloo            & 0.194 & 8.856 & 0.013 \\
        r/Coronavirus          & 0.032 & 9.981 & 0.002 \\
        r/bipolar              & 0.066 & 10.475 & 0.001 \\
        r/CoronavirusCanada    & 0.136 & 9.277 & 0.004 \\
        \midrule
        \textbf{Mean}          & 0.090 & 9.52 & 0.007 \\
        \textbf{Minimum}       & 0.020 & 7.298 & 0.001 \\
        \textbf{Maximum}       & 0.194 & 11.766 & 0.027 \\
        \bottomrule
    \end{tabular}
\end{table}

LDA exhibits KL divergence scores ranging from a minimum of \texttt{0.0199} to a maximum of \texttt{0.1943}, with a mean score of \texttt{0.090}. NMF shows a much broader range of KL divergence scores from \texttt{7.2976} to \texttt{11.7661}, with a mean score of \texttt{9.52}. BERTopic, meanwhile, has a relatively low and consistent range of KL divergence scores, between \texttt{0.0006} and \texttt{0.0265}, with a mean score of \texttt{0.007}. \autoref{tab:reddit_kl_divergence} compares the KL divergence scores of these three topic modelling methods.

 A one-way ANOVA revealed a significant effect of the topic modelling method on KL-Divergence, $F_{2, 33} = 655.30$, $p < .001$, $\eta^2 = .975$. To explore these differences further, a Tukey HSD post hoc test was conducted. The Tukey HSD test indicated that the mean KL-Divergence for LDA ($M = 0.090$) was not significantly higher than BERTopic ($M = 0.007$), with a mean difference of 0.083 (95\% CI, -0.658 to 0.825), $p = 0.95$. Additionally, LDA was significantly lower than NMF ($M = 9.52$), with a mean difference of 9.43 (95\% CI, 8.69 to 10.175), $p < .001$. Furthermore, NMF was significantly higher than BERTopic, with a mean difference of 9.5179 (95\% CI, 8.776 to 10.25), $p < .001$.

\begin{table}[tb!]
    \centering
    \caption[Summary of Results: Execution times]{Execution time in the Reddit datasets using LDA, NMF, and BERTopic (in seconds), lower value is better}
    \label{tab:reddit_execution_time}
    \begin{tabular}{ p{5cm} p{3cm} p{3cm} p{3cm} }
        \toprule
        \textbf{Reddit Dataset} & \textbf{LDA} & \textbf{NMF} & \textbf{BERTopic} \\ 
        \midrule
        r/disabledgamers       & 280 & 51  & 173 \\
        r/MachineLearning      & 1329 & 74  & 701 \\
        r/HermanCainAward      & 3178 & 836 & 2891 \\
        r/cybersecurity        & 933 & 136 & 1190 \\
        r/emergencymedicine    & 452 & 101 & 274 \\
        r/SustainableFashion   & 200 & 53  & 153 \\
        r/LanguageTechnology   & 1034 & 62  & 408 \\
        r/nutrition            & 333 & 68  & 599 \\
        r/uwaterloo            & 531 & 56  & 262 \\
        r/Coronavirus          & 1575 & 91  & 767 \\
        r/bipolar              & 730 & 88  & 653 \\
        r/CoronavirusCanada    & 1587 & 180 & 1178 \\
        \midrule
        \textbf{Mean}          & 1013 & 149 & 771 \\
        \textbf{Minimum}       & 200 & 51  & 153 \\
        \textbf{Maximum}       & 3178 & 836 & 2891 \\
        \bottomrule
    \end{tabular}
\end{table}

\section{Execution Time}\label{time}
For LDA, the execution times range from a minimum of \texttt{200} seconds to a maximum of \texttt{3178} seconds, with a mean time of \texttt{1013} seconds. NMF exhibits a much lower range of execution times from \texttt{51} seconds to \texttt{836} seconds, with a mean time of \texttt{149} seconds. BERTopic shows a range of execution times between \texttt{153} seconds and \texttt{2891} seconds, with a mean time of \texttt{771} seconds. \autoref{tab:reddit_execution_time} compares the execution time of these three topic modelling methods.

A one-way ANOVA revealed a significant effect of the topic modelling method on execution time, $F_{2, 33} = 5.41$, $p = .009$, $\eta^2 = .247$. To explore these differences further, a Tukey HSD post hoc test was conducted. The Tukey HSD test indicated that the mean execution time for LDA ($M = 1013$) was significantly higher than NMF ($M = 149$), with a mean difference of 863.88 (95\% CI, 199.46 to 1528.29), $p = .007$. Additionally, NMF was significantly lower than BERTopic ($M = 771$), with a mean difference of 621.200 (95\% CI, -43.214 to 1285.61), $p = .013$. However, there was no significant difference between LDA and BERTopic, with a mean difference of 242.68 (95\% CI, -421.73 to 907.095), $p = .905$. 

%In summary, NMF demonstrated the lowest execution time compared to both LDA and BERTopic, while no significant differences were observed between LDA and BERTopic.

\subsection{Microsoft Desirability Toolkit }

During the Interview on Topic Models phase, participants were asked to use the Microsoft Desirability Toolkit to select the top 5 words that described each topic modelling technique. Out of 118 words, 64 unique words were chosen by participants: 39 positive (\autoref{tab:positive}), and 25 negative (\autoref{tab:negative}). 

%Our analysis revealed that positive words constitute a significant portion of the total words used, accounting for approximately 63.56\% of the overall feedback. In contrast, negative words make up about 36.44\% of the total words. 

% The CTA Toolkit demonstrates a high frequency of words associated with practical utility and user engagement, such as ``Useful,'' ``Engaging,'' and ``Flexible.'' In contrast, the LDA model emphasizes clarity and relevance, with frequent mentions of ``Clear,'' ``Relevant,'' and ``Easy to use,'' indicating a focus on producing clear and pertinent outputs. NMF is noted for its cleanliness and usability, reflecting its strength in delivering well-organized and user-friendly results, with ``Clean'' and ``Usable'' being the most common descriptors. Lastly, BERTopic highlights usefulness and effectiveness, showcasing its capability to provide valuable and impactful outputs with top positive words like ``Useful,'' ``Meaningful,'' ``Effective,'' ``Valuable,'' and ``Relevant.'' \autoref{tab:positive} presents the frequency of positive words from the Microsoft Desirability Toolkit across the toolkit and the three models.

\begin{table}[tb!]
    \centering
    % \scriptsize
    \caption[Desirability toolkit: Positive words frequency]{Frequency of Positive Words from the Microsoft Desirability Toolkit Across Different Models: LDA, NMF, and BERTopic.}
    \begin{tabular}{l l l}
       \toprule
% & \multicolumn{3}{c}{\textbf{Positive Words}} \\
% \cmidrule{1-3}
\textbf{LDA} & \textbf{NMF} & \textbf{BERTopic} \\
\midrule
Clear (3) & Clean (4) & Useful (3) \\
Relevant (3) & Usable (4) & Meaningful (3) \\
Easy to use (2) & Organized (3) & Effective (3) \\
Useful (2) & Helpful (2) & Valuable (3) \\
Usable (2) & Relevant (2) & Relevant (3) \\
Attractive (1) & Useful (2) & Easy to use (2) \\
Compelling (1) & Straightforward (2) & Satisfying (2) \\
Clean (1) & Expected (2) & Helpful (2) \\
Intuitive (1) & Flexible (1) & Comprehensive (2) \\
Convenient (1) & Connected (1) & Usable (1) \\
Integrated (1) & Comprehensive (1) & Clear (1) \\
Essential (1) & Compatible (1) & Understandable (1) \\
Expected (1) & Accessible (1) & Appealing (1) \\
Advanced (1) & Easy to use (1) & Personal (1) \\
Stimulating (1) & Intuitive (1) & Impressive (1) \\
Innovative (1) & Inspiring (1) & Sophisticated (1) \\
Understandable (1) & Advanced (1) & Engaging (1) \\
Approachable (1) & Stimulating (1) & High quality (1) \\
Organized (1) & Innovative (1) & Advanced (1) \\
Accessible (1) & Consistent (1) & Stimulating (1) \\
Helpful (1) & Inviting (1) & Innovative (1) \\
 & Approachable (1) & Detail (1) \\
 & Understandable (1) & Trustworthy (1) \\
 & Convenient (1) & Unconventional (1)\\
 & Efficient (1) & \\
 & Attractive (1) & \\
 & Compelling (1) & \\
 & Clear (1) & \\
 & Essential (1) & \\
 \\
 \bottomrule
    \end{tabular}
    \label{tab:positive}
\end{table}

% The CTA Toolkit often struggled with balancing detail and usability, sometimes being seen as too simplistic or too complex. Users also found it overwhelming or cluttered, with words like ``Simplistic,'' ``Complex,'' and ``Too technical.'' LDA faced challenges with coherence and depth, often being perceived as simplistic and disconnected. Users also found its outputs lacked clarity and practical value, with the most frequent words being ``Simplistic,'' ``Disconnected,'' and ``Confusing.'' NMF is frequently criticized for lacking sophistication and polish, with issues in managing and ensuring high-quality results. It also faces challenges with usability and flexibility, with the most frequent words being ``Unrefined,'' ``Uncontrollable,'' and ``Poor quality.'' BERTopic is noted for its detailed insights, but users find it difficult to use and integrate. Concerns about trustworthiness and efficiency also arise, indicating it can be overwhelming and hard to visualize if the number of topics is more, with words like ``Complex,'' ``Overwhelming,'' and ``Unrefined.'' \autoref{tab:negative} represents the top negative words from the Microsoft Desirability Toolkit across the toolkit and the three models.

\begin{table}[tb!]
    \centering
    % \scriptsize
    \caption[Desirability toolkit: Negative words frequency]{Frequency of Negative Words from the Microsoft Desirability Toolkit Across Different Models: LDA, NMF, and BERTopic.}
    \begin{tabular}{lll}
       \toprule
\textbf{LDA} & \textbf{NMF} & \textbf{BERTopic} \\
\midrule
Simplistic (5) & Simplistic (5) & Complex (4) \\
Disconnected (4) & Unrefined (3) & Overwhelming (3) \\
Confusing (3) & Uncontrollable (1) & Unrefined (2) \\
Unrefined (2) & Poor quality (1) & Poor quality (1) \\
Not valuable (2) & Busy (1) & Busy (1) \\
Complex (2) & Rigid (1) & Not secure (1) \\
Irrelevant (2) & Complex (1) & Simplistic (1) \\
Limiting (1) & Hard to use (1) & Time-consuming (1) \\
Inconsistent (1) & Confusing (1) & Confusing (1) \\
Undesirable (1) & Time-consuming (1) & Overbearing (1) \\
Rigid (1) & Difficult (1) & Distracting (1) \\
Old (1) & Limiting (1) & Rigid (1) \\
Unpredictable (1) & & Not valuable (1) \\
Poor quality (1) & & Frustrating (1) \\
Difficult (1) & & Impersonal (1) \\
Uncontrollable (1) & & \\
Slow (1) & & \\
\bottomrule
    \end{tabular}
    \label{tab:negative}
\end{table}

\subsection{Qualitative results}  

When asked about incorporating topic modelling into their workflows, 10 out of 12 participants (83\%) were willing and happy to include and/or use it as a supplementary tool in their future work, while two participants (17\%) indicated that their research involves less data, making it less applicable for their needs. This feedback underscores a general sense of enthusiasm towards the techniques and potential for broader adoption. Importantly, they also provided feedback on each model independently, with varying opinions about the suitability of each model for their work. We now summarize these comments.

\subsubsection{LDA}
Researchers provided mixed feedback on the LDA models, appreciating its simplicity and ease of use while acknowledging its limited detail and relevance. A few participants preferred LDA for its familiarity and straightforward results, 3 participants ranking it as their top choice due to ease of access and clarity. LDA was valued for producing clear, organized outputs, making topic keywords accessible for those seeking quick and easy-to-understand results. However, its lack of depth and tendency to generate broad or sometimes irrelevant topics were noted drawbacks.

3 participants perceived LDA as relevant, accessible, and clear. P3 highlighted its relevance and ease of use, stating, ``[For] LDA [I put] `relevant' because I think those topics feel relevant and easy to combine into higher level themes. It's very easy to find information about LDA online and it's pretty easy to learn it.'' P11 appreciated the meaningful groupings created by LDA, noting, ``This model has created groupings that are more meaningful than NMF. There is a common thread across the keywords and there wasn't any random noise blocking the signal of understanding."  They also mentioned the advantage of LDA producing fewer topics, making it easier to focus on each one equally: ``The LDA had the least amount of topics, so it was easier from a user design perspective to pay attention to all of them equally as opposed to the third model [BERTopic] like 56.''  

Participants also noted several limitations of LDA: issues of disconnection between the keywords in the topic clusters, lack of meaningful topics, and confusing clusters. P2 found the topics generated by LDA not semantically meaningful: ``The topics themselves are not really semantically meaningful... if there's just like a word like meaningless, that would describe it.'' P6 felt LDA lacked depth, making it hard to understand topic connections: ``LDA tells you about hey, this is sustainable fashion, but that's it. Some of the words in clusters don't make sense. Like topic six, it has flip flops [and] textiles. I'm not sure how they're relevant.''

\begin{figure}[!h]
    \centering
    \includegraphics[width=1\textwidth]{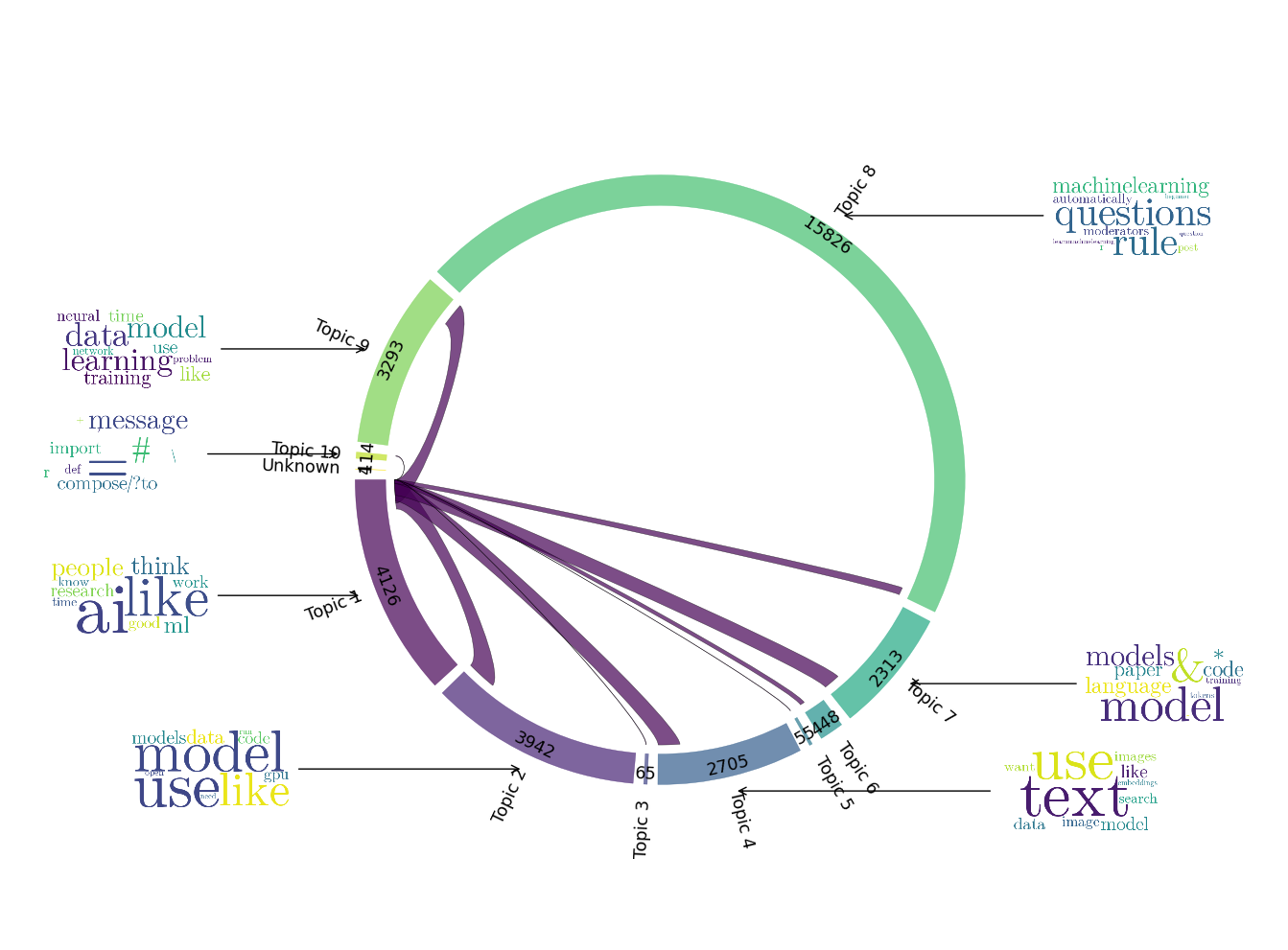}
    \caption[Topic model results for LDA on r/MachineLearning ]{Chord diagram for LDA on the r/MachineLearning dataset, highlighting critiques of irrelevant symbols and the presence of mathematical symbols.}
    \label{fig:lda}
\end{figure}

P7 observed irrelevant content and rogue keywords, leading to unpredictability: ``There was a lot of irrelevant content. There were a bunch of clusters in LDA with rogue keywords. So I wouldn't call it understandable. Unpredictable, I think.'' P8 described LDA as disconnected and rigid, with arbitrarily grouped topics: ``Numerous clustering words could have been in any of the other clusters. It just felt like it was thrown together. Why would increased water intake improve diet quality? Its [clusters] didn't relate to me.'' Additionally, some participants displayed biases towards certain modelling techniques. For instance, P3 stated that ``[I would rank] LDA 1, just because I know it and I'm most familiar with it, I would either choose LDA because I know LDA and that's just what I've always used Or probably bertopic one because it does give me the most option", as displayed in \autoref{fig:lda}.

The default settings for symbol removal in the topic models were not flawless, and LDA was particularly impacted by this issue. This problem led to the inclusion of irrelevant symbols and mathematical characters in the topics, which affected the quality and meaningfulness of the clusters. For instance, P2 indicated that LDA often grouped irrelevant symbols and mathematical characters together, detracting from meaningful clustering: ``They have a topic for just mathematical characters like equal sign and these are all like either code or Math quantifiers or symbols. This one has literally an asterisk. And also like just looking at these, not a lot of them really seem like clusters.'' Another observation from P10 found LDA unrefined, noting the presence of irrelevant symbols and lack of meaningful topics:
``The thing that bugs me the most is like I see sometimes there's emojis. Sometimes there's just like a hyphen or just a space" (Topic keywords: \textit{co, -, op, coop, work, program, job, university, cs, like}).

\subsubsection{NMF}

The NMF models were valued for producing clear, relevant, and organized topics, making them a reliable choice for research. The models generated clean, usable clusters with an organized structure. However, they were sometimes too general, missing nuanced details that other advanced models captured. Overall, NMF balanced detail and usability, offering clear and relevant outputs that were easy to interpret, but fell short in capturing the full complexity of certain topics. Participants valued NMF for its relevance and clear organization, which made it straightforward and usable for understanding current trends. P2 appreciated the succinctness and statistical relevance of NMF.

\begin{figure}[tb]
    \centering
    \includegraphics[width=1\textwidth]{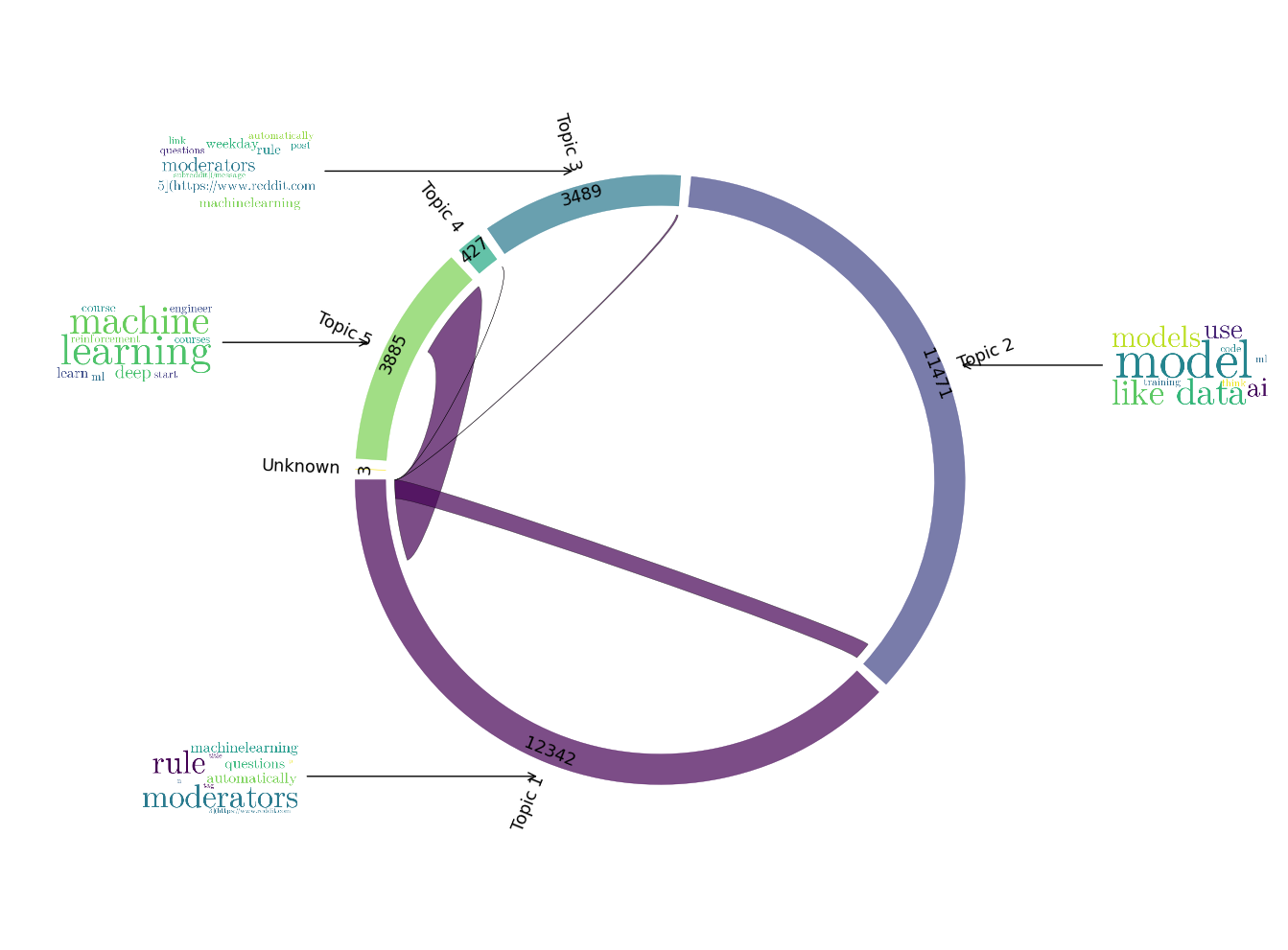}
    \caption[Topic model results for NMF on r/MachineLearning]{Chord diagram for NMF on the r/MachineLearning dataset, highlighting critiques of semantic clarity: topic keywords lacked relevance and required further refinement.}
    \label{fig:nmf}
\end{figure}

Researchers identified several limitations of NMF, highlighting its generality, topic overlap, and lack of semantic clarity. Despite being a useful model, NMF often fell short in providing detailed and relevant topics, particularly for large datasets. P1 mentioned that the number of topics was insufficient, stating, ``This isn't enough. So I think [number of topics as] 5 is definitely not enough for such a large dataset in my opinion.'' P2 further criticized its semantic clarity, noting, ``They [topic keywords] are not semantically meaningful or relevant. It needs a lot more refining at a semantic level", as seen in \autoref{fig:nmf}. This sentiment was echoed by P3, who pointed out the overlap between topics and suggested combining them. Additionally, P7 observed the presence of undesirable keywords and rigidity in the model, emphasizing that ``There were a few clusters with undesirable keywords like cluster three (Topic keywords: \textit{message, compose/?to, \#, import, \&, remindmebot\&amp;subject, ; amp;\#x200b, r)}. It could not capture some topics, so it's a bit rigid in its organization.'' P8 also found NMF too general and lacking specificity, stating, ``NMF is kind of a bit too general. You can say that [some topics] for any university.'' P9 noted that NMF did not capture keywords from other languages as compared to LDA and BERTopic for the r/Coronavirus data, and P11 described NMF as unrefined, confusing, time-consuming, and difficult, concluding, ``Compared to the other two, this one was not pulling together themes relevant to the topic.'' 
  
\subsubsection{BERTopic}

Participants consistently highlighted BERTopic's strengths in producing detailed, organized, and coherent topic clusters. For instance, P1 noted, ``BERTopic had more detailed and organized topics that are usable." Participants felt the topics were clearly separated and logically grouped, which made it easy for them to recognize relevant terms and understand related concepts. P12 highlighted the logical organization of topics, finding the clustering intuitive and easy to comprehend: ``I like how the keywords are captured into 62 topics --- and they seem to make more sense as groupings. But I like that they've broken it out that way because a lot of the [related] language clustered together makes sense.'' 

Additionally, participants felt that BERTopic's topics provided more meaningful and interconnected insights into complex topics than the other two models. For instance, P6 noted, ``Fabrications and dyes are very important topics within the sustainable fashion world because some dyes are more environmentally friendly, and historically, some dyes caused health issues for fashion workers. This is a very important topic that I believe the industry is not really talking a lot about, but it's very interesting to see it here."  This level of detail was seen as crucial for understanding the nuances of specific subjects.  Similarly, they also noted,``The ESG cluster  informs me about investment. I can see the word governance and understand the impact,'' which contrasts with the LDA model's keywords that struggled to effectively link related concepts like ESG and sustainability.

\begin{figure}[tb!]
    \centering
    \includegraphics[width=1\textwidth]{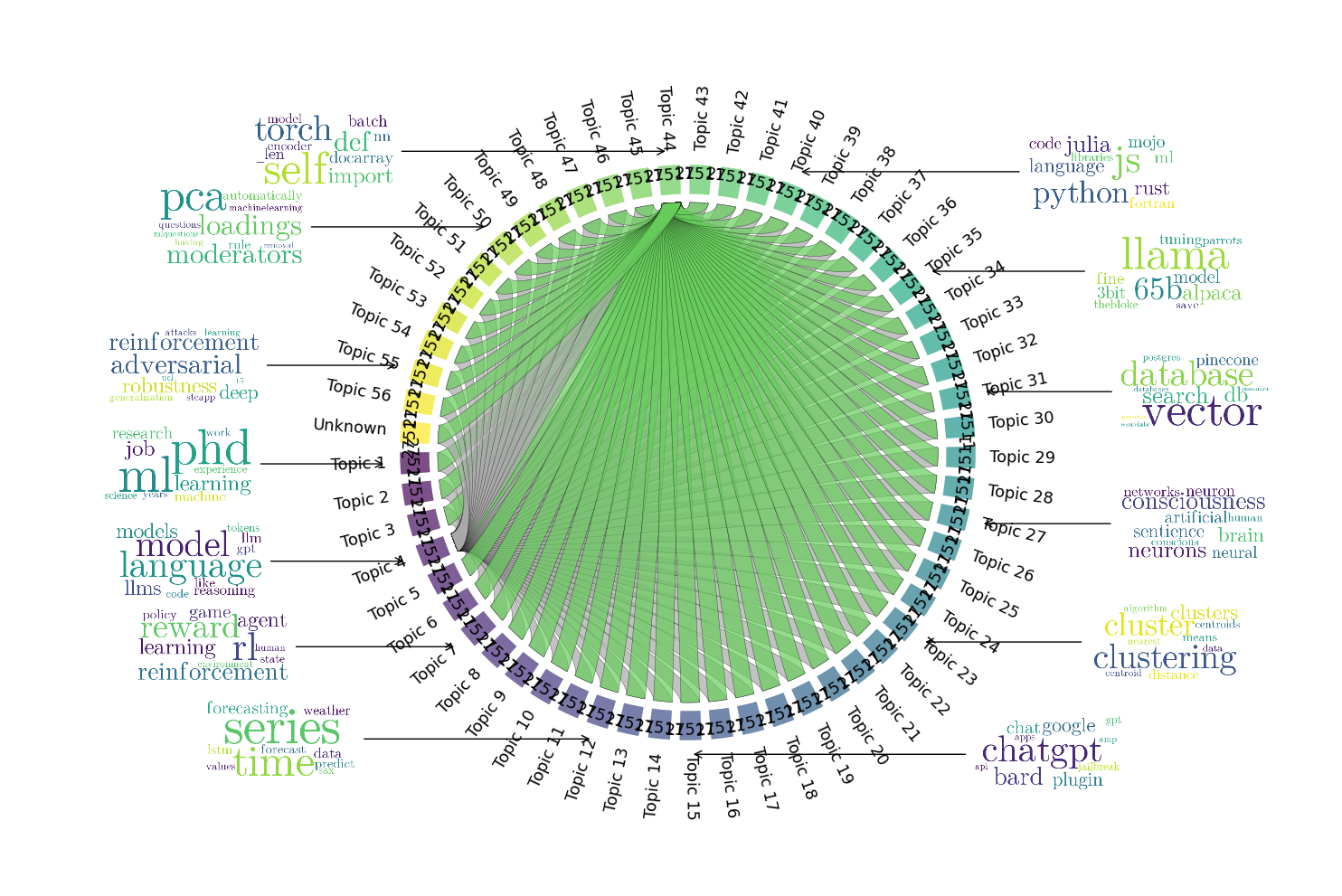}
    \caption[Topic model results for BERTopic on r/MachineLearning]{Chord diagram for BERTopic from the r/MachineLearning dataset. BERTopic's ability to accurately parse and distinguish topics proved especially useful to researchers}
    \label{fig:ml}
\end{figure}

Researchers were impressed by BERTopic's capacity to reveal unexpected yet significant relationships within the data, showcasing its effectiveness in identifying connections that might otherwise be underexplored. P3, initially skeptical about the relevance of certain keywords, found BERTopic intriguing and effective for their analysis. They noted, ``It's very interesting. When I was doing the analysis, I would not have connected horses and dewormer and Apple ivermectin,'' highlighting how BERTopic linked these terms within the context of a surge in demand at farm supply stores due to false claims that the apple-flavored paste could cure or prevent COVID-19. This example underscored BERTopic's capacity to reveal meaningful connections that even the researchers had not anticipated.

% BERTopic's effectiveness in parsing and differentiating topics made it particularly valuable for research, despite taking longer. P2 noted, ``In research, I don't really care how much time I have to spend, but I wanna make sure what I write down on my papers or what I take from a paper has to be relevant. It has to be true."

BERTopic's effectiveness in parsing and differentiating topics made it particularly valuable for research, despite taking longer. P2 emphasized the importance of relevance and accuracy in academic work, stating, ``In research, I don't really care how much time I have to spend, but I wanna make sure what I write down on my papers or what I take from a paper has to be relevant. It has to be true", as seen in \autoref{fig:ml}. Similarly, P9 noted that although BERTopic takes longer, the depth of insights justifies the extra time:``I see the reason and I don't mind staying longer as it's useful for my research." 
P12's request for topics generated ``solely" by BERTopic for their analysis, underscores its superior capacity to deliver comprehensive and actionable insights, reflecting the high level of trust researchers place in its detailed analysis.

However, some researchers found the large number of topics generated by BERTopic overwhelming, which made navigating and digesting the information challenging.  P3 stated,  ``I'm going to say overwhelming as my first one. It is meaningful and a lot of the topics are interesting and relevant, but it is overwhelming and overbearing because I don't think all of those topics are necessary. It's distracting and unrefined."  Due to its complexity, P3 ranked BERTopic lower compared to LDA and NMF. P2 appreciated BERTopic's efficiency but found it slightly overwhelming, suggesting a need for more succinctness, ``I feel like it could be more succinct, like there is room for improvement, but generally I really couldn't complain because as a human, I can just choose to read one or two words." P4 also encountered navigation challenges, finding it difficult to interact with the BERTopic diagram compared to LDA. P11 described BERTopic as unconventional, complex, and busy, although easier to use once understood, but criticized it for being unrefined due to content overload.

% \begin{figure}
%     \centering
%     \includegraphics[width=1\textwidth]{P6/bertopic_Sfashion.png}
%     \caption{BERTopic visualization for r/SustainableFashion}
%     \label{fig:fashion}
% \end{figure}

%======================================================================
\section{Discussion}
%======================================================================

We now discuss the results of our comparative analysis, evaluating the strengths and weaknesses of BERTopic, NMF, and LDA, and exploring their implications for topic modeling. We will examine why BERTopic outperformed other models, the trade-offs involved in using each technique, and the considerations for choosing the most suitable model and evaluation metric for topic modeling.

\subsection{Can LLM-based topic modelling techniques support the thematic analysis workflow?}

We successfully integrated BERTopic into the Computational Thematic Analysis (CTA) Toolkit and received positive feedback from participants. For instance, our interviews and hands-on evaluations indicated that BERTopic's contextually rich and coherent topics were highly valuable to researchers. P7 stated, ``The clusters [in BERTopic] are more fine-grained and easier to understand." In turn, the detailed and coherent topic clusters provided a deeper understanding of intricate datasets. For instance, P3 highlighted the importance of BERTopic in uncovering semantic relationships between ``horses," ``dewormer," and ``apple ivermectin" in their COVID-19 data, and P6 showed how BERTopic revealed connections between sustainable fashion topics like the relevance of ``fabrications" and ``dyes" in environmental contexts. 

These findings contrast recent work in the HCI literature which suggests that qualitative researchers may be hesitant to adopt computational techniques in their work \cite{feuston2021putting,jiang2021supporting}. Where in the past researchers have cited concerns about intimacy, ownership, and agency in their own analyses, our findings show how LLM-based techniques might \emph{enhance} their ability to understand data, to reveal new trends, and prompt qualitative researchers to explore new aspects of collected data. Thus, integrating LLM-based unsupervised topic modelling techniques into qualitative thematic analysis workflows is not only feasible but can also be highly beneficial.  We argue that future work should focus on understanding how computational methods can support qualitative researchers, while maintaining their agency in the process.

%Our study demonstrated that BERTopic also can uncover significant relationships in the data of value to researchers, revealing connections in the data which might not be that evident during data pre-processing. Several studies have highlighted the potential of LLM-based techniques in enhancing social media data analysis. Notably, research by \citet{egger2022topic,abuzayed2021bert} shows that models like BERTopic excel in capturing semantic nuances and contextual information. 

%Previous studies have shown that LLMs excel in understanding context and subtle themes, providing a nuanced and detailed categorization of topics, which surpasses traditional methods \citep{wu2023large, tang2023evaluating}. 

\subsection{How does BERTopic compare to LDA and NMF?}

When comparing BERTopic to conventional unsupervised modelling techniques, participants consistently highlighted its strengths in producing detailed, organized, and coherent topic clusters. They appreciated how BERTopic clearly separated and logically grouped topics, making it easier to recognize relevant terms and understand related concepts. Participants consistently praised BERTopic for being ``comprehensive",``satisfying,'' and ``sophisticated'', underscoring its capabilities. Further, our quantitative results showed that BERTopic consistently produced higher coherence scores, greater topic diversity, and more meaningful, relevant topics when compared to LDA. Indeed, 8 out of 12 researchers rated BERTopic as their preferred model. 

In contrast, we found that LDA and NMF often produce topics with mixed themes, making them difficult to interpret~\citep{egger2022topic}. While LDA and NMF have been widely used for topic modelling in qualitative data analysis, our study suggests that it may be beneficial to reconsider this choice. Moreover, our results help us understand how quantitative measures might be used to improve topic models to support qualitative analysis. While measures like coherence have traditionally been used to quantitatively evaluate topic model performance, there has been criticism within the HCI community around this practice \cite{gauthier2022will,baumer2017comparing}. That is, in qualitative research models provide ``scaffolding for human interpretation'' \cite{baumer2017comparing} and there is a risk of over-emphasizing the need to find \emph{optimal} models rather than models that support qualitative research.

Notably, if we had optimized for coherence, in many cases we would have selected NMF over BERTopic. Similarly, NMF often provided high scores for topic diversity and the lowest execution time of all techniques. Instead we assert that participants preferred models that provided a high number of rich and distinct topics. BERTopic consistently provided the highest number of topics for each data set, while maintaining high topic diversity scores. In turn, participants consistently cited the specific and detailed topics as being what they found most useful in interpreting the data.

%Our analysis also demonstrates that human researchers read and consider all topics in context \citep{li2023can}. For instance, P6 highlighted that BERTopic's topics were in-depth, such as Topic 7 (Topic Keywords: \textit{leather, vegan, veganism, plant, plastic, animal, based, products, diet, meat}), which explores leather extensively, including vegan leather and plastic. Additionally, P2 appreciated BERTopic's ability to identify trending topics in AI, specifically llama-based chat GPT variants: ``These are all talking specifically about Llama-based chat GPT variants. So that's quite fine-grained and nice to see" (Topic Keywords: \textit{llama, llama2, cpp, mayonnaise, ms, xethub, ggml, file, run, mount}). 
%This demonstrates that BERTopic performed better in capturing a broader spectrum of subtopics as compared to LDA and NMF. This broader and more detailed approach is crucial because it allows for a more comprehensive understanding of the data, providing organized, coherent, and detailed insights that can significantly enhance the depth and breadth of qualitative research analysis.

\subsection{What are the needs and preferences of qualitative researchers and how can these be addressed through unsupervised topic modelling?}

We identified three general needs among qualitative researchers: 

%the ability of a topic model to perform optimal data cleaning, if execution time is a priority, and the optimal number of topics. 

% Additionally, while determining the ideal number of topics is challenging with LDA and NMF \citep{saqib2019analysis}, our model employed data-driven, built in ``auto" generation techniques to suggest optimal topic numbers, effectively addressing this issue. 

%We will discuss our findings related to each of these needs. Firstly, we discuss the tradeoff of data cleaning in traditional and LLM based approaches. Secondly, researchers praised BERTopic's effectiveness in identifying trends within social media data. Thirdly, we assessed whether execution time is a priority for researchers when choosing a topic modellingtool. Lastly, we examined the optimal number of topics required to provide meaningful insights without overwhelming the analysis.

\subsubsection{Data Cleaning}

The primary tradeoff in using LDA and NMF is the necessity for extensive data cleaning and preprocessing, which requires significant manual effort. This labour-intensive process, involving the removal of irrelevant symbols, stemming, and lemmatizing can be time-consuming and prone to human error \citep{gauthier2022computational,gauthier2020}. In contrast, modern techniques like BERTopic are designed to streamline data cleaning and processing, potentially identifying and grouping overlooked keywords with less human intervention \citep{sawant2022enhanced,an2023marketing}. This approach aims to save time, improve consistency, and allow researchers to focus more on interpreting and applying insights. BERTopic’s ability to process raw data and produce coherent topic clusters with reduced preprocessing highlights its efficiency and robustness, leading to more efficient and insightful qualitative research by balancing human expertise.

But this raises a critical question: should we continue with manual data cleaning, or automate the process using advanced LLM-based topic modelling techniques? The benefits of automating data cleaning with BERTopic include reduced manual effort, minimized errors, and enhanced consistency, making it a compelling choice for modern qualitative research \citep{grootendorst2022bertopic}. However, it is essential to consider whether the automated process can fully capture the nuances that human expertise brings to data interpretation \citep{vayansky2020review,owoahene2024review}. Future research should explore the balance between automation and manual intervention to optimize both efficiency and depth of insight in topic modelling \citep{holzinger2016interactive,yang2021multimodal}.

\subsubsection{Execution Time is not a Priority}

Participants valued BERTopic's granularity and prioritized quality over speed in their research. As detailed in the \autoref{time}, LDA took significantly longer than NMF, and NMF took significantly less time than BERTopic, with no significant difference in execution time between LDA and BERTopic. It is important to acknowledge that participants did not create the models themselves and therefore did not need to consider processing time unless explicitly discussed. Despite NMF's quicker processing times, many participants were willing to wait longer for BERTopic's results due to its superior granularity and detailed insights. For instance, P9 noted, ``I see the reason and I don’t mind staying longer as it’s useful for my research." This suggests that while efficiency is important, the quality and depth of the results are essential for researchers, even if it means longer processing times. BERTopic outperformed LDA, the current ``gold standard," across all quantitative measures in our study, underscoring its significant potential and relevance in the field.

%Overall, while BERTopic is highly valued for its nuanced and relevant topic clusters, addressing  usability challenges like visualization would enhance its utility, making it a more balanced and accessible modellingtechnique.

\subsubsection{Navigating BERTopic's Complexity for Deeper Insights}

BERTopic's complexity, while initially challenging, ultimately provides substantial efficacy in terms of insight and depth. As users become familiar with its structure, the model's capacity to reveal nuanced patterns and relationships becomes evident, highlighting its potential for comprehensive analysis. This underscores the importance of user engagement and learning in maximizing the benefits of advanced topic modelling techniques like BERTopic, leading to more informed and meaningful research outcomes.

In our study, the methods like Hierarchical Edge Bundles \citep{pujara2012large,dou2013hierarchicaltopics}, Interactive Topic Maps \citep{hu2014interactive}, and Force-Directed Layouts \citep{van2008visualizing} could have better facilitated navigation and interpretation of complex topic relationships by grouping the increased number of topics into hierarchical structures; thus reducing the overwhelming and complex visualization of BERTopic.  

Furthermore, BERTopic can incorporate hierarchical topic modeling, addressing the limitations of flat topic clusters seen in LDA \citep{grootendorst2022bertopic}. Visualizing the hierarchy enables researchers to gain a deeper understanding of the thematic organization within their corpus, facilitating more nuanced topic analysis. The lack of hierarchical visualization is a severe limitation of the CTA toolkit, as it restricts researchers from fully exploring and interacting with complex topic models. Our future research would aim to develop tools that provide capabilities for visualizing and interacting with hierarchical structures, allowing for a more comprehensive analysis and understanding of the data.

\section{Limitations}

Our study also uncovered several important limitations that highlight areas for improvement. Researchers highlighted the need for simplified interfaces and better visualizations, such as interactive chord diagrams and hierarchical representations, to manage large datasets. They also suggested future features, including a more in-depth search function, to further improve the design and guide the next phase of our project.

\textbf{Optimizing LDA and NMF Models:} Future research should focus on optimizing parameters and evaluation metrics, especially the criteria for determining the number of topics. While topic coherence proved to be a valuable metric, relying solely on it for model evaluation posed challenges. This was evident when we used coherence to determine the number of topics for LDA and NMF, potentially limiting their performance and applicability.  The presence of irrelevant symbols and mathematical characters in the topics highlighted a need for more precise and automated data cleaning processes. Both LDA and NMF required extensive manual effort to clean the data accurately. 

\textbf{BERTopic's Number of Topics:} Participants often found the large number of topics generated by BERTopic to be overwhelming, which led some to prefer other models despite acknowledging BERTopic's superior detail and depth. Exploring hierarchical topic models may provide more organized and interpretable results. However, those models would necessitate improved visualization techniques. Developing more intuitive and effective visual interfaces will help users better understand and interact with the topic models.

\textbf{Generalizability:} Ensuring the generalizability and practical utility of our findings requires a larger, more diverse participant base and longitudinal studies. Our study included a diverse group of researchers, but the sample size was relatively small, with only 12 participants. This limitation restricts the generalizability of our findings. A larger and more varied participant base would help validate and extend our conclusions, ensuring that the results are applicable across a broader spectrum of qualitative research contexts. Further, conducting longitudinal studies where researchers use the software over extended periods of time can provide deeper insights into its practical utility and areas for improvement.

%======================================================================
\section{Conclusion and Future Work}
%======================================================================
Integrating BERTopic into the Computational Thematic Analysis (CTA) Toolkit significantly enhanced its capabilities. Researchers praised BERTopic for generating detailed, coherent topics, their ease of interpretation, and their utility in uncovering hidden relationships within data. Evaluation metrics showed BERTopic’s superiority in topic coherence and topic diversity. This enhancement improved qualitative data analysis efficiency and depth, combining advanced computational methods with an intuitive interface for comprehensive and logical data representation.

In reflecting on our interviews, we found that researchers prefer models that produce detailed, organized, and coherent topic clusters, making it easier to recognize relevant terms and understand related concepts. They value models that reveal unexpected yet significant relationships within the data, providing comprehensive and actionable insights. The ability to capture a wide range of nuances and generate a larger number of topics is particularly important for deep understanding of complex subjects. Our results demonstrate the potential LLM-based methods like BERTopic for supporting qualitative analysis of large data sets.

\bibliographystyle{ACM-Reference-Format}
\bibliography{uw-ethesis} % references.bib is the bibliography file

\end{document}